\documentclass[letter]{aa} 

\usepackage{graphicx}
\usepackage{txfonts}
\usepackage[utf8]{inputenc}
\begin{document} 

   \title{Accretion versus core-filament collision
}
   \subtitle{Implications for streamer formation in Per-emb-2}

   \author{Fumitaka Nakamura\inst{1,2,3},
          Quang Nguyen-Luong\inst{4,5,6},
          Kousuke Ishihara\inst{1,2},
          Aoto Yoshino\inst{1,3}
}
   \institute{
   National Astronomical Observatory of Japan, National Institutes of Natural Sciences, 2-21-1 Osawa, Mitaka, Tokyo 181-8588, Japan
   \label{NAOJ}
   \and
    Department of Astronomical Science, SOKENDAI (The Graduate University for Advanced Studies), 2-21-1 Osawa, Mitaka, Tokyo 181-8588, Japan
    \label{SOKENDAI}
    \and
    Department of Astronomy, Graduate School of Science, The University of Tokyo, 7-3-1 Hongo, Bunkyo-ku, Tokyo 113-0033, Japan
    \label{TODAI}
    \and
   Astrophysical Data Sciences, CSMES, the American University of Paris, PL111, 2 bis, passage Landrieu, 75007 Paris, France  
   \label{AUP}
   \and
    Universit\'e Paris-Saclay, Universit\'e Paris Cit\'e, CEA, CNRS, AIM, 91191, Gif-sur-Yvette, France
  \label{CEA}
  \and
TNU Observatory, Tay Nguyen University, 567 L\'e Duan, Ea Tam, Bu\^{o}n Ma Thu\^{o}t, Đ\u{a}k L\u{a}k, 630000, Vietnam
\label{TNU}
  \and \email{fumitaka.nakamura@nao.ac.jp}
             }

   \date{\today}

 
  \abstract
   {
Recent millimetre and sub-millimetre observations have unveiled
 elongated and asymmetric structures around protostars. These structures, referred to as streamers, often exhibit coherent velocity gradients, seemingly indicating a directed gas flow towards the protostars. However, their origin and role in star formation remain uncertain.
}
   {
   The protostellar core Per-emb-2, located in Barnard 1, has a relatively large streamer of $10^4$ au that is more prominent in emission from carbon-chain molecules. We aim to unveil the formation mechanism of this streamer. 
}
   {
 We conducted mapping observations towards Per-emb-2 using the Nobeyama 45 m telescope. We targeted carbon-chain molecular lines such as CCS, HC$_3$N, and HC$_5$N at  45 GHz.
}
{
Using \texttt{astrodendro}, we identified one protostellar and four starless cores, including three new detections, on the {\it Herschel} column density map. The starless and protostellar cores are more or less gravitationally bound. 
We discovered strong CCS and HC$_3$N emissions extending from the north to the south, appearing to bridge the gap between the protostellar core and the starless core to its north. This bridge spans $3\times 10^4$ au with velocities of 6.5 to 7.0 km s$^{-1}$.
The velocity gradient  of the bridge is opposite that of the streamer. 
Thus, the streamer is unlikely to be connected to the bridge, suggesting that the former does not have an accretion origin.
}
   {
   We propose that a collision between a spherical core and the filament has shaped the density structure in this region, consequently triggering star formation within the head-tail-shaped core.
In this core-filament collision scenario, the collision appears to have fragmented the filament into two structures. 
The streamer is a bow structure, while the bridge is a remnant of the shock-compressed filament.
Thus, we conclude that the Per-emb-2 streamer does not significantly contribute to the mass accumulation towards the protostar.}
\keywords{stars: formation - ISM: clouds - ISM: structure - submillimeter: ISM}

\maketitle

\titlerunning{Accretion vs Core Filament Collision}
\authorrunning{Nakamura, F., Nguyen-Luong, Q., Ishihara, K., \& Yoshino, A.}
%
\section{Introduction}
\label{sec:intro}

\begin{figure*}[!hbpt]
 \begin{center}
 \includegraphics[angle=0, totalheight=6cm]
 {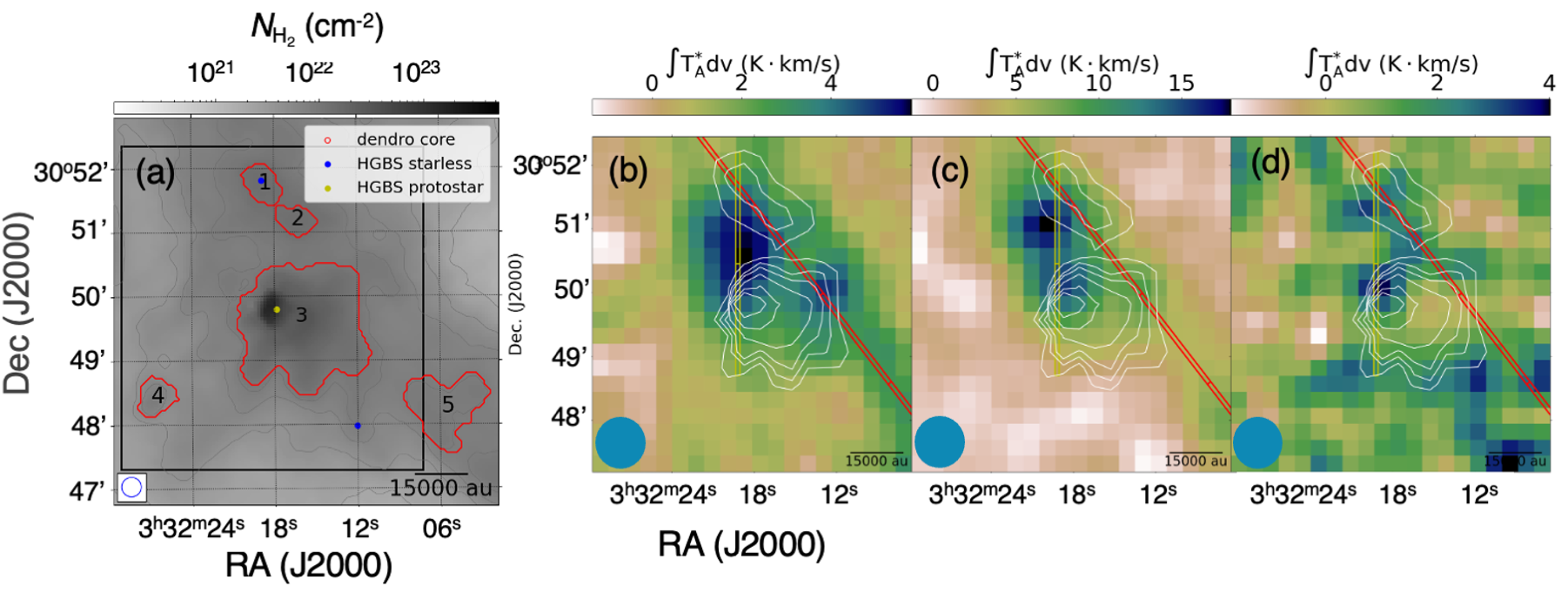}\hspace{1cm}\\
 \end{center}
\caption{
{\bf Overall distributions of molecular gas. Panel (a):} Observed area as seen on the {\it Herschel} hydrogen molecule column density map. The black box is the 5\arcmin $\times$ 5\arcmin observation box. The five enclosed red contours indicate the dense cores identified by \texttt{astrodendro} (see Table \ref{tab:dendo_core1} for the core numbers).
The yellow dot located at the centre is the position of the protostar. 
The blue dots show the positions of the starless cores identified by \citet{pezzuto21}. 
\textbf{Panels (b)-(d):} Velocity-integrated maps (from 6.0 to 8.5 km s$^{-1}$) of CCS ($J_N=4_3-3_2$) {\bf (b)}, HC$_3$N ($J=5-4$) {\bf (c)}, and HC$_5$N ($J=17-16$) {\bf (d)}  observed with the NRO 45~m telescope. 
The red lines follow the large-scale filament in this area. The filament is more prominent in a map that covers a larger area (see Fig. \ref{fig:largemap}).
The blue ellipses in panels (b), (c), and (d) are the map effective beam size.
}
\label{fig:obsarea}
\end{figure*}

Stars form from dense cores in molecular clouds. 
According to the standard scenario of star formation \citep{shu87}, a nearly spherical dense core undergoes gravitational contraction, forming a nearly axi-symmetric system consisting of a protostar and a rotationally supported disk. 
However, recent observations have uncovered significant non-axisymmetric structures surrounding young protostellar systems, commonly referred to as streamers or spirals \citep{pineda20, sanhueza19, chen21,valdivia22,hsieh23,olguin23}. 
Hereafter, we use the term `streamer' to refer to such structures for the sake of simplicity.  These structures often exhibit coherent velocity patterns, interpreted as infall towards the central protostar \citep{pineda20, sanhueza21}. Understanding the role of streamers in mass loading towards central stars is crucial for unveiling the complexities of the stellar mass accumulation process.

Numerical studies have proposed various mechanisms for streamer formation. For example, in many numerical simulations of cluster formation, intermediate-density structures connected to dense cores and sink particles, which resemble the streamers, are often observed, and these structures play a role in mass accretion onto the central protostellar system  \citep[e.g.][]{bonnell06,nakamura07,kuffmeier17,padoan20}.   
On the other hand, \citet{tu2023} demonstrate that streamer-like structures could arise from local turbulent compression within magnetised parent cores, without the need for external accretion of ambient gas. In this case, the protostellar system does not need to have structures connected to the inter-core medium,
and streamers can be generated through the accretion process within the natal core.
Therefore, the accretion of ambient gas likely plays only a minor role in supplying the gas, and the stellar mass is roughly determined by the parent core mass.  
\citet{yano24} propose that streamers could form through dense core collisions \citep[DCCs; see also][]{kinoshita22}. They estimated the frequencies of DCCs in nearby star-forming molecular clouds and found that the typical dense core experiences at least one collision in its lifetime, particularly in clustered environments.
In this DCC model, the compressed layer created between two cores evolves into streamers and spirals, creating a rotating envelope around the protostellar system.  In this case, 
the streamer supplies additional mass to the central protostar, but the mass accretion occurs only temporarily and depends on conditions such as the impact parameter and collision speed.
However, observational evidence is needed to further elucidate these mechanisms.

To shed light on streamer formation, we conducted observations towards Per-emb-2, a core with a streamer towards the central protostar \citep{pineda20}. 
It is located in the southern part of Barnard 1 in the Perseus molecular cloud at a distance of 300 pc from the Sun \citep{zucker19}. 
Gas within the streamer appears to exhibit a velocity gradient towards the central protostar -- from 7.5 km s$^{-1}$ (far side) to 7.0 km s$^{-1}$  (protostar) -- and the streamer is particularly prominent in emissions from chemically early-phase molecules such as CCS and HC$_3$N, as observed with the Northern Extended Millimetre Array (NOEMA) interferometer.
Since the abundance of these molecules decreases steeply  with time in high-density environments, 
\citet{pineda20} claim that 
the gas of the streamer is supplied from outside via accretion.
Very recently, \citet{taniguchi24} suggested that the ambient gas may be provided from the northern part of the protostar.

We present detailed observational information in Sect.~\ref{sec:obs} and the analysis of the velocity and spatial distributions of molecular gas obtained from these observations  
in Sect. \ref{sec:result}.
In Sect. \ref{sec:discussion} we propose our core-filament collision (CFC) model, which explains the structure in the observed region.

\section{Observations and data}
\label{sec:obs}

\begin{figure*}[htbp]
 \begin{center}
 \includegraphics[angle=0, width=0.85 \textwidth]{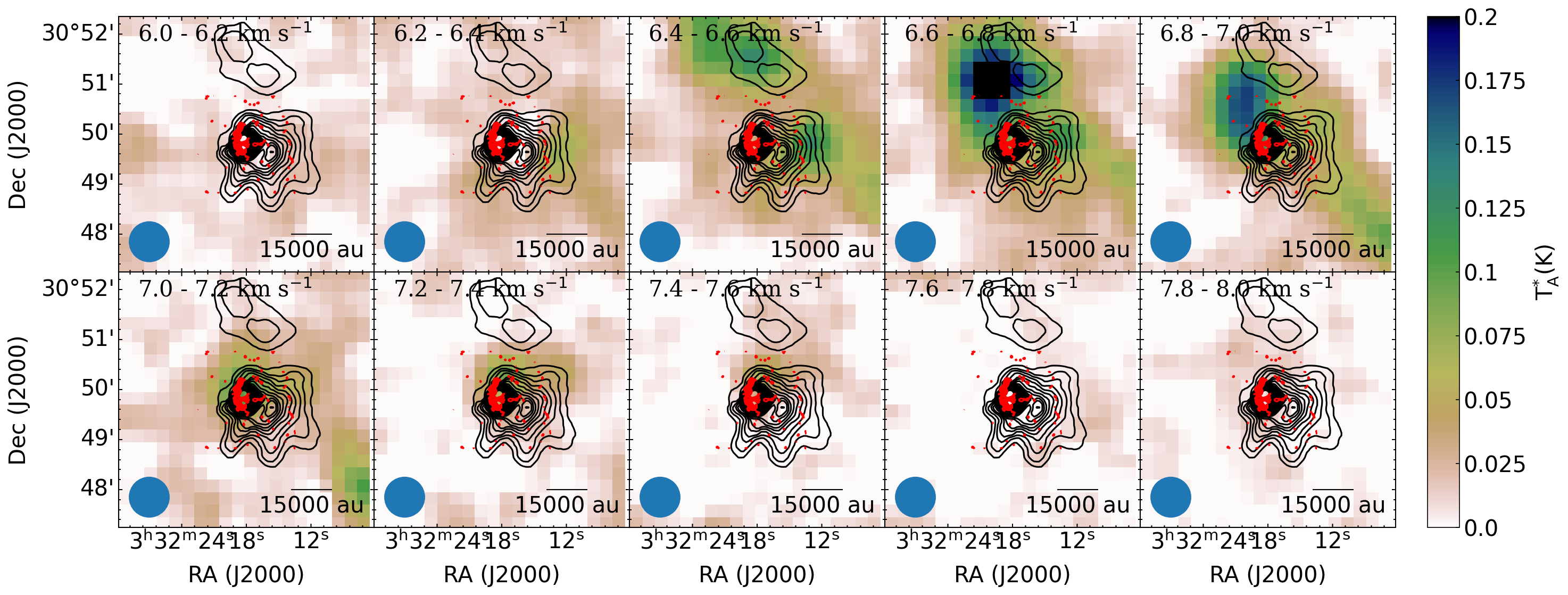}
 \end{center}
\caption{Colour-scale image showing channel maps of CCS ($J_N=4_3-3_2$). The red contours are the CCS ($J_N=8_7-7_6$) integrated intensity contours from NOEMA obtained by \cite{pineda20}. The black contours are {\it Herschel} column densities derived by \cite{pezzuto21}, from $1\times10^{22}$ to $10\times10^{22}$~cm$^{-2}$ in steps of $0.2\times10^{22}$~cm$^{-2}$. The blue ellipse is the map effective beam size.
}
\label{fig:CCSchannelmaps}
\end{figure*}

We carried out the on-the-fly mapping observations towards the protostellar core Per-emb-2 using the Z45 receiver of the Nobeyama Radio Observatory (NRO) 45m telescope \citep{nakamura14} in the 2022--2023 season. Details of the observations are presented in Appendix \ref{sec:obs+line}.
The mapped area was determined such that the protostar is located at the centre of the $5' \times 5'$ box
(Fig. \ref{fig:obsarea}a).

The target lines are CCS ($J_N=4_3-3_2$), 
 HC$_3$N ($J=5-4$), and HC$_5$N ($J=17-16$). The parameters are summarised in Table \ref{tab:targetline}.
See \citet{nakamura24} for the accurate rest frequency of CCS ($J_N=4_3-3_2$).  
The telescope has a full width at half maximum beam size of 38\arcsec \, at 43 GHz.
The standard chopper wheel method was used to convert the output signal to the antenna temperatures ($T_{A}^*$) 
and to correct for the atmospheric attenuation.
The main beam efficiency of Z45 is $\eta_{\rm Z45}\simeq 0.73$ at 45 GHz.
The main-beam temperature was calculated as $T_{\rm mb} = T_{A}^*/\eta_{\rm Z45}$.
We adopted a spheroidal function as a gridding convolution function to calculate the intensity at each grid point of the final cube data with a spatial grid size of 15\arcsec \, and a frequency resolution of 3.81 kHz, which corresponds to $\sim 0.025$ km s$^{-1}$ at 45 GHz. 
The final effective angular resolution of the map was 49\arcsec, corresponding to 15000 au, or 0.07 pc at the distance of 300 pc 
\citep{zucker19}. 

We downloaded the high-resolution H$_2$ column density and dust temperature maps obtained by the {\it Herschel} Gould Belt Survey \citep{andre10,pezzuto21}.
The angular resolution of the maps is 18\arcsec.

\begin{center}
\hspace*{-1cm}
 \begin{table*}
 \footnotesize
  \caption{Properties of cores detected in the mapping region by \texttt{astrodendro}.}
 \begin{tabular}{cccccccccll}
 \hline
 Number & Peak R.A. & Peak Dec. & $M_{\rm core}$ & $R_{\rm core}$ & $\left<N\right>$  & $\left<n\right>$ & $\left<T_{\rm dust}\right>$ & $\sigma_{\rm CCS}$ & $\alpha_{\rm vir} (\alpha_{\rm BE}$) & type / HGBS \\
 & (J2000) & (J2000) & $M_\odot$ & pc & $10^{22}$ cm$^{-2}$  & $10^5$ cm$^{-3}$ & K & km s$^{-1}$ &  &  \\
 \hline
 1 & 3:32:19.13 & 30:51:44.67 & 0.508 & 0.0148 & 1.28  & 5.4 & 12.7 & 0.37 & 1.422 (0.565) & starless/449 \\
 2 & 3:32:16.36 & 30:51:08.40 & 0.398 & 0.0077 & 1.25  & 29.6 & 12.6 & 0.41 & 1.014 (0.377) & starless/new \\
 3 & 3:32:18.50 & 30:49:50.63 & 9.723 & 0.0884 & 1.97  & 0.5 & 12.1 & 0.84 & 1.077 (0.176) & protostellar/447 \\
 4 & 3:32:26.94 & 30:48:21.45 & 0.200 & 0.0130 & 0.64  & 3.1 & 13.8 & -- & -- (1.265) & starless/new \\
 5 & 3:32:06.46 & 30:47:55.46 & 1.083 & 0.0565 & 0.86  & 0.2 & 12.7 & 0.45\tablefootmark{a} & 2.954 (1.013) & starless/new \\
 \hline
 HGBS 442  & 3:32:12.04& 30:47:59.0 & 0.105 & 0.0300  & 0.25 & 0.05 & 0.02 &  10.4 & (3.9) & starless\\ 
 \hline
 \end{tabular}
\label{tab:dendo_core1}
\tablefoot{The core numbers are from the {\it Herschel} core catalogue of \citet{pezzuto21}.}
\tablefoottext{a}{We estimated this velocity dispersion only within the area overlapping with the observation box.}
\end{table*}
\end{center}

\section{Results and analysis}
\label{sec:result}

\subsection{Dense core identification and their dynamical states}

In Fig. \ref{fig:obsarea}a we present the {\it Herschel} column density map of the observed area.
We applied {\texttt{astrodendro} \citep{rosolowsky08} to this image and identified five cores, four of which are fully within our observation box (see Appendix \ref{subsec:core idenification} for more details). The boundaries of the identified cores are outlined in red. Their physical quantities are summarised in Table \ref{tab:dendo_core1}.
In the same area, \citet{pezzuto21} identified three cores using getsources: a protostellar core (HGBS 447) and two starless cores (HGBSs 442 and 449).
Our cores 1 and 3 correspond to HGBS 442 and 447, respectively.  Cores 2, 4, and 5 are starless and newly detected. We could not identify HGBS 449, which has the lowest column density.
The central protostellar core (core 3) has a mass of $\sim 10 M_\odot$.  According to \citet{pineda20}, the protostellar core has a systemic velocity of $\sim~7.0$ km s$^{-1}$ and has a streamer with a length of $10^4$ au \citep{pineda20}.

From the CCS cube data, we derived the core's CCS velocity dispersion ($\sigma_{\rm CCS}$) 
and then converted it to the intrinsic velocity dispersion ($\sigma_{\rm tot}$) of gas with a mean molecular weight of $2.33 m_H$ to assess its dynamical state (see Appendix \ref{sec:dynamical state} for the details of the analysis).
Cores 1, 2, 3, and 5 have $\alpha_{\rm vir}\sim 1$; the virial parameter $\alpha _{\rm vir}$ was computed under the assumption of a centrally condensed ($\propto r^{-2}$) sphere. The protostellar core has a significantly small Bonnor-Ebert ratio of $\alpha_{\rm BE}\sim$ 0.18, but its $\alpha_{\rm vir}$ is close to unity (the Bonnor-Ebert ratio is the ratio between the Bonnor-Ebert critical mass of the core and the observed mass).
Core 4 does not have significant CCS emission, and we could not measure the velocity dispersion, but its Bonnor-Ebert ratio ($\alpha_{\rm BE}\sim$ 1.3) suggests it is close to bound. 
We also calculated the external pressures exerted on the core surfaces.\ We find that they play a minor role except for core 5, for which we calculated the core's external pressure ($3 \rho_{\rm amb} \sigma_{\rm tot, amb}^2$) from the ambient density ($\rho_{\rm amb}$) and the ambient gas intrinsic velocity dispersion ($ \sigma_{\rm tot, amb}$) measured in the branch just below the corresponding core. The branch is an intermediate structure identified by \texttt{astrodendro}.
Core 5 is dynamically compressed by the ambient pressure and likely did not form via spontaneous gravitational fragmentation, a process that leads to cores becoming self-gravitating.
The other cores are close to being in dynamical equilibrium.

For the protostellar core, the column density distribution is not symmetric: it exhibits a head-tail shape. The centre of gravity determined from this distribution is significantly offset from the position of the protostar itself (see Fig. \ref{fig:largemap}). 
This suggests that the streamer does not point directly towards the gravity centre of the protostellar core.
If its curved structure is generated by the angular momentum of the infalling gas, a large rotating envelope ($\sim 10^3$ au) should form around the protostar and connect to the streamers.
However, there is currently no observational evidence of such a large structure \citep[e.g.][]{tobin16, pineda20}. 
Therefore, it is unlikely that the streamer is driven by gravitational infall from outside the core.

\begin{figure*}[htbp]
 \begin{center}
 \includegraphics[angle=0, width=0.8\textwidth]
 {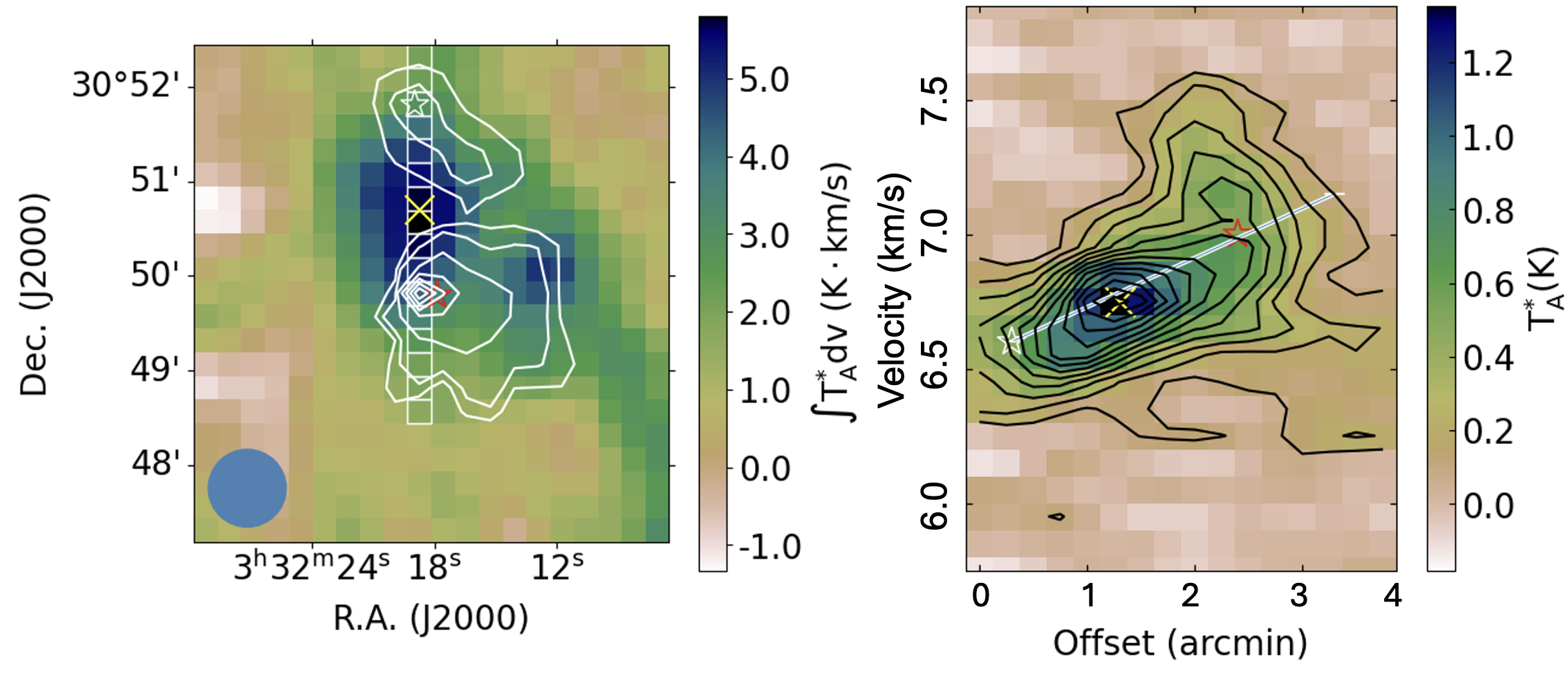}
 \end{center}
\caption{Velocity structures in Per-emb-2. {\bf Left:} Integrated-intensity map of CCS (from 6.0 to 8.0 km\,s$^{-1}$) in colour overlaid with the {\it Herschel} column density map in contours (as in Fig.~\ref{fig:obsarea}) and the position of the velocity cut along the bridge (in white). {\bf Right:} Position-velocity plots along the bridge structure from the cut along the bridge seen on the left. White and red stars indicate the positions of starless cores and the protostellar core. The yellow cross indicates the position of the CCS peak. The blue ellipse is the map effective beam size.
}
\label{fig:pv}
\end{figure*}

\subsection{CCS/HC$_3$N bridge with a length of $3\times 10^4$ au (0.2 pc)
}
In Fig. \ref{fig:obsarea} we present the velocity-integrated maps of CCS, HC$_3$N, and HC$_5$N alongside the {\it Herschel} column density map.
In the CCS integrated intensity map, we recognise the two prominent CCS elongated features: (1) an elongated feature, stretching from north to south and  
bridging the gap between the starless cores 1 and 2 and the protostellar core 3, and (2) a compact emission  positioned to the west of the protostellar core.

The distribution of HC$_3$N emission appears to resemble that of CCS, with strong concentrations at the southern periphery of the starless cores 1 and 2. Another significant peak is evident immediately above the protostar,
 and a fainter peak is located to the west, coinciding with a local maximum in CCS intensity. 
HC$_5$N, which exhibits the weakest intensity of the three lines, has a peak at a similar position as the second peak of HC$_3$N, or the northern portion of the protostellar core.

Figures \ref{fig:CCSchannelmaps} and \ref{fig:channelmap} display the velocity channel maps of CCS,  HC$_5$N, and HC$_3$N. The starless cores 1 and 2 appear to be associated with relatively strong compact emission around a velocity of 6.5 km s$^{-1}$.

The CCS emission within the velocity range 6.5--6.7 km s$^{-1}$ exhibits a fragmented filamentary structure, extending along a line from the north-east to the south-west that coincides with the large-scale filament seen in the H$_2$ column density image
(see Fig. \ref{fig:largemap}). An elongated structure is discernible in the south-east corner of the CCS map, though the peak is faint in the H$_2$ map (see Fig. \ref{fig:obsarea}).

At a velocity of 6.6--7.0 km s$^{-1}$, the elongated structure observed in CCS gradually connects with the central protostellar core, forming a bridge. A similar bridging structure is also observed in the corresponding velocity channel for HC$_5$N.
Overall, these emissions demonstrate a weak velocity gradient that spans from 6.5 km s$^{-1}$ to 7.0 km s$^{-1}$ towards the protostar.

\subsection{Mismatch in velocity structure between the bridge and streamer
}
From the observations detailed in the previous section, it becomes evident that the protostellar core 3 and the northern starless cores 1 and 2 exhibit a connection characterised by a bridge structure.
In Fig. \ref{fig:pv} we present a CCS position-velocity plot that traces the 0.2 pc bridge structure.

The protostellar core 3 exhibits a velocity of $\sim$ 7 km s$^{-1}$, indicating a velocity disparity between the two cores bridged by an intermediate-velocity region. 
The velocity gradient along the bridge is estimated to be 
$\approx $ 0.5 km s$^{-1}$/0.2 pc $\sim$ 2.5 km s$^{-1}$ pc$^{-1}$.
The existence of the intermediate-velocity gas bridging the two structures with different velocities suggests a dynamical interaction between the cores, a feature similar to larger-scale cloud-cloud collisions \citep[e.g.][]{nakamura12a,nguyen13,fukui21,kinoshita21b}. The enhanced CCS emission observed could potentially stem from the colliding region, where CCS formation may be accelerated due to increased density. Both CCS and HC$_3$N exhibit critical densities of around $10^4$ cm$^{-3}$, implying that while the shocked layer may have a higher density, the column density might not be sufficiently high to be recognised in the 2D {\it Herschel} image.

The streamer seemingly follows the $\sim 10^{22}$ cm$^{-2}$ contour of the protostellar core and 
spatially overlaps with the distribution of the CCS emission (see Fig. \ref{fig:CCSchannelmaps}), but the velocities are very different. 
The streamer has a much steeper velocity gradient in the opposite direction to that of the bridge (see the right panel of Fig. \ref{fig:pv}).
Therefore, we conclude that the bridge is not a structure connected to the streamer.

\begin{figure}[thbp]
 \begin{center}
  \includegraphics[angle=0, width= 0.5 \textwidth]{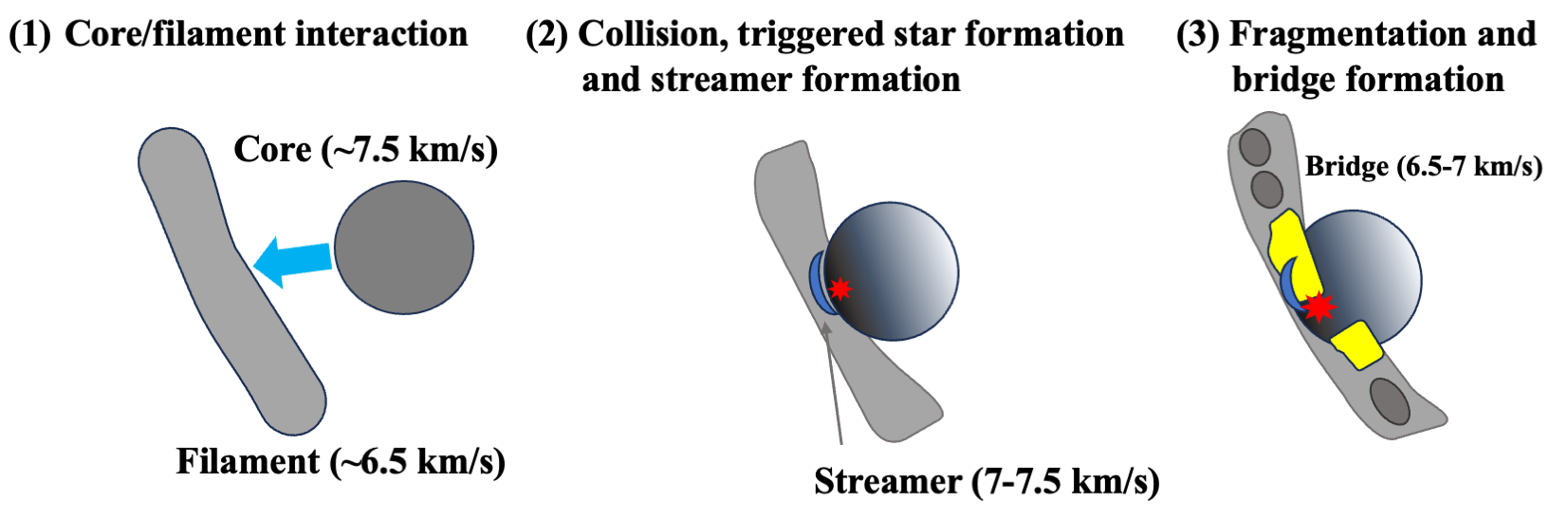}
 \end{center}
\caption{CFC model of Per-emb-2. (1) A dense core is approaching a filament (grey) with a speed 
of $\sim$ 1 km s$^{-1}$. (2) They collide, protostar formation is triggered at the shock-compressed part,
and a streamer (dark blue) of size $3\times 10^4$ au is formed. (3) The dense core shreds the filament into two unequal parts, and cores 1, 2, and 5 are created. The shock-compressed parts of the filament become a bridge (yellow), where the formation of carbon-chain molecules is accelerated by the increased density. 
}
\label{fig:model}
\end{figure}

\section{Discussion}
\label{sec:discussion}

\subsection{Accreting streamer or not?}

\citet{taniguchi24} claim, based on the integrated intensity map, that the streamer connects to the northern region through the bridge. They suggest that the streamer is an accreting flow. However, 
as mentioned above, the gas in the bridge is likely a separate component since the streamer and bridge have different velocities: the velocity of the bridge ranges from 6.5 (in the north) to 7.0 km s$^{-1}$ (south) with 
a velocity gradient of 2.5 km s$^{-1}$ pc$^{-1}$ (hereafter, we define this direction as positive), whereas the streamer has velocities of 7.5 (north) to 7.0 km s$^{-1}$ (south) with 
a velocity gradient of --0.5 km s$^{-1}$/0.05 pc = --10 km s$^{-1}$ pc$^{-1}$.
 From on the velocity distribution, there is no evidence that the streamer connects to the bridge.

In addition, the centre of gravity of the protostellar core is significantly offset from the protostar, by $\sim$ 12\arcsec ($\sim 7000$ au). If the streamer originates from the inflow from the ambient gas driven by the gravity, it likely points towards the core's gravity centre  (see Fig. \ref{fig:largemap}).
If the offset were due to rotation, this object would have a large rotating envelope with a size of $\sim 10^3$ au. Such a structure is not detected towards the protostar \citep{tobin16}. This also implies that the streamer is not an origin of accretion, contrasting with the interpretation of \citet{pineda20}.
The northern cores (1 and 2) are close to gravitationally bound and therefore could potentially evolve into protostars. Their merging could trigger star formation. 
The mass of the protostellar core is significant; a part of it may eventually accrete onto the central protostar before the ambient gas falls in.

\citet{yano24} used the DCC scenario to reproduce the density structure in Per-emb-2 in numerical simulations.  According to this model, when two cores collide with each other, a shock-compressed layer is formed in between them and connects to the protostar. The gas is inflowing towards the protostar.  However, 
if this is the case, both the bridge and streamer should have velocity gradients in the same direction \citep[see][]{kinoshita22}. 
The velocity and density structures appear more complicated than what the DCC model predicts.
Therefore, the formation mechanism of the streamer is neither ambient gas inflow nor two-round core collision.

\subsection{A core-filament collision (CFC) model}
Based on the evidence of the existence of the large-scale filament and a positive gradient towards the streamer of the protostar, we propose a CFC model to explain the configuration of Per-emb-2.\ Figure \ref{fig:model} depicts the schematic representation of this CFC model. 

We propose that: (1) A spherical core, with a relative speed of $\sim$ 1--1.5 km s$^{-1}$, approached from the west side of the filament and collided with it. (2) The collision initiated star formation on the far side of the eastern extremity of the core, where the initial shock compression occurred, leading to the formation of a protostellar system.
The arc-like streamer may be the bow structure formed by the CFC, and therefore the velocity gradient of the streamer is not due to gravitational accretion. 
If its dynamics are not significantly affected by the stellar gravity, the flow along the streamer should be pointed towards the outside  (i.e. it is an outflow and not infalling).
The bow typically has an almost linear velocity structure because of its curvature. Such a feature is consistent with the observation. 
When the stellar gravity is sufficiently strong to attract the gas towards the star, the flow may be pointing towards the protostar.
Since it is difficult to judge the direction (outflow or inflow) from the current observational data, numerical experiments are needed to verify this scenario more quantitatively.

The head-tail morphology observed in core 3 agrees well with what would result from such a collision.
The CCS arc (streamer) structure \citep{pineda20} appears to trace the edge of the protostellar core ($\sim 1-2\times 10^{22}$ cm$^{-2}$; see Fig. {\ref{fig:CCSchannelmaps}}). 

Furthermore, the collision led to the division of the filament into two distinct structures (cores 1 and 2 and core 5). 
The shock-compressed layers in the northern and southern regions attained densities of $ \sim 10^4$ cm$^{-3}$, resulting in relatively strong CCS and HC$_3$N emissions.
The northern shock-compressed layer became interconnected with the central protostellar system. There is a possibility that it could evolve into an accreting streamer in the future. However, the total accretion mass via the bridge (future streamer) may be low. The protostellar core envelope still contains enough material ($\sim 10 M_\odot$). 
Therefore, the accretion along the streamers, if it exits, plays only a minor role in the mass accumulation towards the central protostar.

The total velocity difference is about 1 (= 7.5--6.5)
km s$^{-1}$, and the filament width is about 0.1 pc; therefore, it takes 0.1 pc / 1 km s$^{-1}$  $\sim 10^5$ yr for the core to cross the filament.
This is comparable to the typical lifetime of a Class 0 protostar and
 is consistent with the triggered star formation proposed in the CFC model scenario.
Furthermore, if the bridge is the remnant of the shock-compressed layer, its density must have increased from $n_{\rm amb} \sim 10^3$ cm$^{-3}$ to $\sim 10^4$ cm$^{-3}$ due to the isothermal shock with a Mach number of 1.5 km s$^{-1}$/sound speed $\sim 6$.
This condition in the shocked area, or streamer, is similar to the condition assumed by \citet{suzuki92}, who started the chemical evolution calculation assuming a steady state with a uniform isothermal gas of $10^4$ cm$^{-3}$ and $T=10$ K.
According to Fig. 13 of \cite{suzuki92}, 
CCS and HC$_3$N become most abundant in $\sim 10^5$ yr in such a density range, and declines steeply after 0.5 Myr \citep[see also][]{leung84}.
These timescales match the CFC scenario.
In our CFC or DCC scenarios, shocks are likely to increase the gas temperature somewhat, but the molecular gas is highly radiative and therefore the gas temperature stays low even after shock compression. 
In such a situation, the streamers formed by the collision model may have emission of SO or CH$_3$OH, whose sublimation temperatures are about 50 and 80 K, respectively. 
These lines could exhibit broader widths as a result of shock interactions. This hypothesis should be validated with chemical evolution calculations using, for example, the Paris-Durham shock code \citep{godard19}.

\begin{acknowledgements}

This work was financially supported by JSPS KAKENHI Grant Numbers JP23H01218 (F.N.).
Part of this work was supported by the NAOJ Visiting Research Grant (Q.N.L.).
We thank the NRO staff for both operating the 45 m and helping us with the data reduction.
We thank the anonymous referee for valuable comments that helped improve the paper.
\end{acknowledgements}

%
%

\bibliographystyle{aa} 
\bibliography{nakamura} 

\begin{thebibliography}{32}
\expandafter\ifx\csname natexlab\endcsname\relax\def\natexlab#1{#1}\fi

\bibitem[{{Andr{\'e}} {et~al.}(2010){Andr{\'e}}, {Men'shchikov}, {Bontemps},
  {K{\"o}nyves}, {Motte}, {Schneider}, {Didelon}, {Minier}, {Saraceno},
  {Ward-Thompson}, {di Francesco}, {White}, {Molinari}, {Testi}, {Abergel},
  {Griffin}, {Henning}, {Royer}, {Mer{\'{\i}}n}, {Vavrek}, {Attard},
  {Arzoumanian}, {Wilson}, {Ade}, {Aussel}, {Baluteau}, {Benedettini},
  {Bernard}, {Blommaert}, {Cambr{\'e}sy}, {Cox}, {di Giorgio}, {Hargrave},
  {Hennemann}, {Huang}, {Kirk}, {Krause}, {Launhardt}, {Leeks}, {Le Pennec},
  {Li}, {Martin}, {Maury}, {Olofsson}, {Omont}, {Peretto}, {Pezzuto}, {Prusti},
  {Roussel}, {Russeil}, {Sauvage}, {Sibthorpe}, {Sicilia-Aguilar}, {Spinoglio},
  {Waelkens}, {Woodcraft}, \& {Zavagno}}]{andre10}
{Andr{\'e}}, P., {Men'shchikov}, A., {Bontemps}, S., {et~al.} 2010, \aap, 518,
  L102

\bibitem[{{Bertoldi} \& {McKee}(1992)}]{bertoldi92}
{Bertoldi}, F. \& {McKee}, C.~F. 1992, \apj, 395, 140

\bibitem[{{Bonnell} \& {Bate}(2006)}]{bonnell06}
{Bonnell}, I.~A. \& {Bate}, M.~R. 2006, \mnras, 370, 488

\bibitem[{{Chen} {et~al.}(2021){Chen}, {Ren}, {Li}, {Liu}, {Wang}, {Shen},
  {Ellingsen}, {Sobolev}, {Mei}, {Li}, {Wu}, \& {Kim}}]{chen21}
{Chen}, X., {Ren}, Z.-Y., {Li}, D.-L., {et~al.} 2021, \apjl, 923, L20

\bibitem[{{Fukui} {et~al.}(2021){Fukui}, {Habe}, {Inoue}, {Enokiya}, \&
  {Tachihara}}]{fukui21}
{Fukui}, Y., {Habe}, A., {Inoue}, T., {Enokiya}, R., \& {Tachihara}, K. 2021,
  \pasj, 73, S1

\bibitem[{{Godard} {et~al.}(2019){Godard}, {Pineau des For{\^e}ts}, {Lesaffre},
  {Lehmann}, {Gusdorf}, \& {Falgarone}}]{godard19}
{Godard}, B., {Pineau des For{\^e}ts}, G., {Lesaffre}, P., {et~al.} 2019, \aap,
  622, A100

\bibitem[{{Hsieh} {et~al.}(2023){Hsieh}, {Segura-Cox}, {Pineda}, {Caselli},
  {Bouscasse}, {Neri}, {Lopez-Sepulcre}, {Valdivia-Mena}, {Maureira},
  {Henning}, {Smirnov-Pinchukov}, {Semenov}, {M{\"o}ller}, {Cunningham},
  {Fuente}, {Marino}, {Dutrey}, {Tafalla}, {Chapillon}, {Ceccarelli}, \&
  {Zhao}}]{hsieh23}
{Hsieh}, T.~H., {Segura-Cox}, D.~M., {Pineda}, J.~E., {et~al.} 2023, \aap, 669,
  A137

\bibitem[{{Kinoshita} \& {Nakamura}(2022)}]{kinoshita22}
{Kinoshita}, S.~W. \& {Nakamura}, F. 2022, \apj, 937, 69

\bibitem[{{Kinoshita} {et~al.}(2021){Kinoshita}, {Nakamura}, {Nguyen-Luong},
  {Wu}, {Shimoikura}, {Sugitani}, {Dobashi}, {Takemura}, {Sanhueza}, {Kim},
  {Kang}, {Evans}, {White}, \& {Fallscheer}}]{kinoshita21b}
{Kinoshita}, S.~W., {Nakamura}, F., {Nguyen-Luong}, Q., {et~al.} 2021, \pasj,
  73, S300

\bibitem[{{Kuffmeier} {et~al.}(2017){Kuffmeier}, {Haugb{\o}lle}, \&
  {Nordlund}}]{kuffmeier17}
{Kuffmeier}, M., {Haugb{\o}lle}, T., \& {Nordlund}, {\r{A}}. 2017, \apj, 846, 7

\bibitem[{{Leung} {et~al.}(1984){Leung}, {Herbst}, \& {Huebner}}]{leung84}
{Leung}, C.~M., {Herbst}, E., \& {Huebner}, W.~F. 1984, \apjs, 56, 231

\bibitem[{{Maruta} {et~al.}(2010){Maruta}, {Nakamura}, {Nishi}, {Ikeda}, \&
  {Kitamura}}]{maruta10}
{Maruta}, H., {Nakamura}, F., {Nishi}, R., {Ikeda}, N., \& {Kitamura}, Y. 2010,
  \apj, 714, 680

\bibitem[{{Nakamura} {et~al.}(2024){Nakamura}, {Chiong}, {Taniguchi}, {Chien},
  {Ho}, {Hwang}, {Yeh}, {Shimoikura}, {Yamasaki}, {Liu}, {Hirano}, {Lai},
  {Nishimura}, {Kawabe}, {Dobashi}, {Fujii}, {Yonekura}, {Ogawa}, \& {Nguyen
  Luong}}]{nakamura24}
{Nakamura}, F., {Chiong}, C.-C., {Taniguchi}, K., {et~al.} 2024, \pasj, 76, 563

\bibitem[{{Nakamura} \& {Li}(2007)}]{nakamura07}
{Nakamura}, F. \& {Li}, Z.-Y. 2007, \apj, 662, 395

\bibitem[{{Nakamura} {et~al.}(2012){Nakamura}, {Miura}, {Kitamura},
  {Shimajiri}, {Kawabe}, {Akashi}, {Ikeda}, {Tsukagoshi}, {Momose}, {Nishi}, \&
  {Li}}]{nakamura12a}
{Nakamura}, F., {Miura}, T., {Kitamura}, Y., {et~al.} 2012, \apj, 746, 25

\bibitem[{{Nakamura} {et~al.}(2014){Nakamura}, {Sugitani}, {Tanaka},
  {Nishitani}, {Dobashi}, {Shimoikura}, {Shimajiri}, {Kawabe}, {Yonekura},
  {Mizuno}, {Kimura}, {Tokuda}, {Kozu}, {Okada}, {Hasegawa}, {Ogawa}, {Kameno},
  {Shinnaga}, {Momose}, {Nakajima}, {Onishi}, {Maezawa}, {Hirota}, {Takano},
  {Iono}, {Kuno}, \& {Yamamoto}}]{nakamura14}
{Nakamura}, F., {Sugitani}, K., {Tanaka}, T., {et~al.} 2014, \apjl, 791, L23

\bibitem[{{Nguyen-Lu'o'ng} {et~al.}(2013){Nguyen-Lu'o'ng}, {Motte}, {Carlhoff},
  {Louvet}, {Lesaffre}, {Schilke}, {Hill}, {Hennemann}, {Gusdorf}, {Didelon},
  {Schneider}, {Bontemps}, {Duarte-Cabral}, {Menten}, {Martin}, {Wyrowski},
  {Bendo}, {Roussel}, {Bernard}, {Bronfman}, {Henning}, {Kramer}, \&
  {Heitsch}}]{nguyen13}
{Nguyen-Lu'o'ng}, Q., {Motte}, F., {Carlhoff}, P., {et~al.} 2013, \apj, 775, 88

\bibitem[{{Olguin} {et~al.}(2023){Olguin}, {Sanhueza}, {Chen}, {Lu}, {Oya},
  {Zhang}, {Ginsburg}, {Taniguchi}, {Li}, {Morii}, {Sakai}, \&
  {Nakamura}}]{olguin23}
{Olguin}, F.~A., {Sanhueza}, P., {Chen}, H.-R.~V., {et~al.} 2023, \apjl, 959,
  L31

\bibitem[{{Padoan} {et~al.}(2020){Padoan}, {Pan}, {Juvela}, {Haugb{\o}lle}, \&
  {Nordlund}}]{padoan20}
{Padoan}, P., {Pan}, L., {Juvela}, M., {Haugb{\o}lle}, T., \& {Nordlund},
  {\r{A}}. 2020, \apj, 900, 82

\bibitem[{{Pezzuto} {et~al.}(2021){Pezzuto}, {Benedettini}, {Di Francesco},
  {Palmeirim}, {Sadavoy}, {Schisano}, {Li Causi}, {Andr{\'e}}, {Arzoumanian},
  {Bernard}, {Bontemps}, {Elia}, {Fiorellino}, {Kirk}, {K{\"o}nyves},
  {Ladjelate}, {Men'shchikov}, {Motte}, {Piccotti}, {Schneider}, {Spinoglio},
  {Ward-Thompson}, \& {Wilson}}]{pezzuto21}
{Pezzuto}, S., {Benedettini}, M., {Di Francesco}, J., {et~al.} 2021, \aap, 645,
  A55

\bibitem[{{Pineda} {et~al.}(2020){Pineda}, {Segura-Cox}, {Caselli},
  {Cunningham}, {Zhao}, {Schmiedeke}, {Maureira}, \& {Neri}}]{pineda20}
{Pineda}, J.~E., {Segura-Cox}, D., {Caselli}, P., {et~al.} 2020, Nature
  Astronomy, 4, 1158

\bibitem[{{Rosolowsky} {et~al.}(2008){Rosolowsky}, {Pineda}, {Kauffmann}, \&
  {Goodman}}]{rosolowsky08}
{Rosolowsky}, E.~W., {Pineda}, J.~E., {Kauffmann}, J., \& {Goodman}, A.~A.
  2008, \apj, 679, 1338

\bibitem[{{Sanhueza} {et~al.}(2019){Sanhueza}, {Contreras}, {Wu}, {Jackson},
  {Guzm{\'a}n}, {Zhang}, {Li}, {Lu}, {Silva}, {Izumi}, {Liu}, {Miura},
  {Tatematsu}, {Sakai}, {Beuther}, {Garay}, {Ohashi}, {Saito}, {Nakamura},
  {Saigo}, {Veena}, {Nguyen-Luong}, \& {Tafoya}}]{sanhueza19}
{Sanhueza}, P., {Contreras}, Y., {Wu}, B., {et~al.} 2019, \apj, 886, 102

\bibitem[{{Sanhueza} {et~al.}(2021){Sanhueza}, {Girart}, {Padovani}, {Galli},
  {Hull}, {Zhang}, {Cortes}, {Stephens}, {Fern{\'a}ndez-L{\'o}pez}, {Jackson},
  {Frau}, {Kock}, {Wu}, {Zapata}, {Olguin}, {Lu}, {Silva}, {Tang}, {Sakai},
  {Guzm{\'a}n}, {Tatematsu}, {Nakamura}, \& {Chen}}]{sanhueza21}
{Sanhueza}, P., {Girart}, J.~M., {Padovani}, M., {et~al.} 2021, \apjl, 915, L10

\bibitem[{{Shu} {et~al.}(1987){Shu}, {Adams}, \& {Lizano}}]{shu87}
{Shu}, F.~H., {Adams}, F.~C., \& {Lizano}, S. 1987, \araa, 25, 23

\bibitem[{{Suzuki} {et~al.}(1992){Suzuki}, {Yamamoto}, {Ohishi}, {Kaifu},
  {Ishikawa}, {Hirahara}, \& {Takano}}]{suzuki92}
{Suzuki}, H., {Yamamoto}, S., {Ohishi}, M., {et~al.} 1992, \apj, 392, 551

\bibitem[{{Taniguchi} {et~al.}(2024){Taniguchi}, {Pineda}, {Caselli},
  {Shimoikura}, {Friesen}, {Segura-Cox}, \& {Schmiedeke}}]{taniguchi24}
{Taniguchi}, K., {Pineda}, J.~E., {Caselli}, P., {et~al.} 2024, \apj, 965, 162

\bibitem[{{Tobin} {et~al.}(2016){Tobin}, {Looney}, {Li}, {Chandler}, {Dunham},
  {Segura-Cox}, {Sadavoy}, {Melis}, {Harris}, {Kratter}, \& {Perez}}]{tobin16}
{Tobin}, J.~J., {Looney}, L.~W., {Li}, Z.-Y., {et~al.} 2016, \apj, 818, 73

\bibitem[{{Tu} {et~al.}(2024){Tu}, {Li}, {Lam}, {Tomida}, \& {Hsu}}]{tu2023}
{Tu}, Y., {Li}, Z.-Y., {Lam}, K.~H., {Tomida}, K., \& {Hsu}, C.-Y. 2024,
  \mnras, 527, 10131

\bibitem[{{Valdivia-Mena} {et~al.}(2022){Valdivia-Mena}, {Pineda},
  {Segura-Cox}, {Caselli}, {Neri}, {L{\'o}pez-Sepulcre}, {Cunningham},
  {Bouscasse}, {Semenov}, {Henning}, {Pi{\'e}tu}, {Chapillon}, {Dutrey},
  {Fuente}, {Guilloteau}, {Hsieh}, {Jim{\'e}nez-Serra}, {Marino}, {Maureira},
  {Smirnov-Pinchukov}, {Tafalla}, \& {Zhao}}]{valdivia22}
{Valdivia-Mena}, M.~T., {Pineda}, J.~E., {Segura-Cox}, D.~M., {et~al.} 2022,
  \aap, 667, A12

\bibitem[{{Yano} {et~al.}(2024){Yano}, {Nakamura}, \& {Kinoshita}}]{yano24}
{Yano}, Y., {Nakamura}, F., \& {Kinoshita}, S.~W. 2024, \apj, 964, 119

\bibitem[{{Zucker} {et~al.}(2019){Zucker}, {Speagle}, {Schlafly}, {Green},
  {Finkbeiner}, {Goodman}, \& {Alves}}]{zucker19}
{Zucker}, C., {Speagle}, J.~S., {Schlafly}, E.~F., {et~al.} 2019, \apj, 879,
  125

\end{thebibliography}

\begin{appendix} 

\section{Observations and target lines}
\label{sec:obs+line}

The on-the-fly observations were carried out in October 2022 (2 nights) and in March 2023 (1 daytime). Z45 is dual linear polarisation receiver and we summed up the horizontal and vertical polarisation components to improve the signal-to-noise ratios.
The scan interval of the on-the-fly observations was set to 8\arcsec, about a fifth of the beam size.  The pointing observations were made every one hour with SiO maser lines of NML-Tau (KL Tau). The pointing errors were within 3\arcsec -- 5\arcsec.
We used SAM45 digital spectrometer as a backend. SAM45 is a highly flexible FX-type digital spectrometer with a finest frequency resolution of 3.81 kHz. The bandwidth of SAM45 with 3.81 kHz was 15.625 MHz.
It took about 30 min to obtain a single map. We summed 6 maps to obtain a final map.
The total observation time was about 6 hours including the overhead of pointing observations. The on-source time was 3 hours.
During the observations, the typical system noise temperatures were about 150--200 K.

The rest frequencies and transitions of the target lines are summarised in Table \ref{tab:targetline} (see \citealt{nakamura24} for the rest frequency of CCS).

\begin{center}
 \begin{table}[h]
  \caption{Target lines.}
 \begin{tabular}{ccc}
 \hline
 Species & Transition & Frequency\\ \hline
 - & - & (GHz)\\ \hline
CCS  & $J_N=4_3-3_2$ & 45.379033 \\ 
HC$_3$N & $J=5-4$ & 45.490316 \\
HC$_5$N & $J=17-16$ & 45.264720 \\
   \hline
 \end{tabular}
 \label{tab:targetline}
\end{table}
\end{center}

\section{Core identification and physical quantities
}
\subsection{Core identification}
\label{subsec:core idenification}

We applied \texttt{astrodendro} \citep{rosolowsky08} to the high-resolution {\it Herschel} column density map with an angular resolution of 18\arcsec.
The \texttt{astrodendro} has three parameters for the hierarchical structure analysis: \texttt{min\_value} which is the threshold value,  \texttt{min\_delta} which is the minimum step for structure identification, and  \texttt{min\_npix} which is the minimum pixel number.
We adopt \texttt{min\_value} $ = 3\sigma $, \texttt{min\_delta} $=1.5 \sigma $, and \texttt{min\_npix} = beam area/pixel size, $\sigma \ (= 6.81 \times 10^{20}$ cm$^{-2}$) is the rms noise level.
The \texttt{astrodendro} identifies three structures: a leaf, branch, and trunk. 
The minimum structure of `a leaf' is defined as a dense core.
To estimate the ambient gas pressure around the cores, we adopted the physical quantities averaged in a branch immediately below the corresponding leaf.
We also derive the core mass by subtracting the background component which is defined as the trunk.

\subsection{Dynamical states of the cores}
\label{sec:dynamical state}

We assess the dynamical states of the cores applying the virial theorem.
The virial equation for a spherical core is given by
\begin{equation}
\frac{1}{2}\frac{\partial ^2 I}{\partial t^2}
 = U+ W+ S \ ,
\end{equation}
where $I$, $U$, $W$, and $S$ are the moment of inertia, internal kinetic energy, gravitational energy, and surface pressure term, respectively, and the magnetic field terms are neglected \citep{maruta10}.
These individual terms are given as
\begin{eqnarray}
    U &=& 3 M_{\rm core} \sigma_{\rm tot}^2  \\
    W &=& -\frac{3}{5}a \frac{GM_{\rm core}^2}{R_{\rm core}}  \\
    S &=& -4 \pi R^3_{\rm core} P_{\rm ex}
,\end{eqnarray}
where $G$, $M_{\rm core}$, $R_{\rm core}$, and $P_{\rm ex}$ are the gravitational constant, core mass, core radius, and the external pressure, respectively.
$a$ is a dimensionless parameter of order unity which measures the effects of a non-uniform or nonspherical mass distribution 
\citep{bertoldi92}. For a uniform sphere and a centrally condensed sphere with $\rho \propto r^{-2}$, $a$ = 1 and 5/3, respectively. Here, we adopt $a=5/3$ since the cores tend to be more or less centrally-condensed.
$\sigma_{\rm tot}$ is the 1D intrinsic velocity dispersion of the molecule of mean mass,
\begin{equation}
    \sigma_{\rm tot} = \sqrt{\sigma_{\rm CCS} ^2+ k_B 
    T \left( \frac{1}{\mu m_H} - \frac{1}{m_{\rm CCS}} \right)
    }
    \label{eq:total velocity dispersion}
,\end{equation}
and $G$ is the gravitational constant, $k_B$ is the Boltzmann constant, $\sigma_{\rm CCS}$ is the velocity dispersion measured by CCS, $\mu$ (=2.33) is the mean molecular weight, $m_H$ is the mass of a hydrogen atom, and $m_{\rm CCS}$ is the mass of a CCS molecule. 
The CCS velocity dispersions towards the cores are computed by the Gaussian fitting of the averaged spectra 
(see Fig. \ref{fig:spec}).
It is worth noting that CCS emission may preferentially trace the outer envelopes of cores, given its critical density of $10^4-10^5$ cm$^{-3}$. In contrast, the core densities calculated from the {\it Herschel} image range from $3\times 10^6$ to 
$ 2\times 10^4$ cm$^{-3}$ as shown in Table \ref{tab:dendo_core1}. Despite this, in the core analysis presented here, we adopt the CCS line widths as representative of the cores' properties.

\begin{figure*}
\begin{tabular}{c}
\includegraphics[width=18.cm]{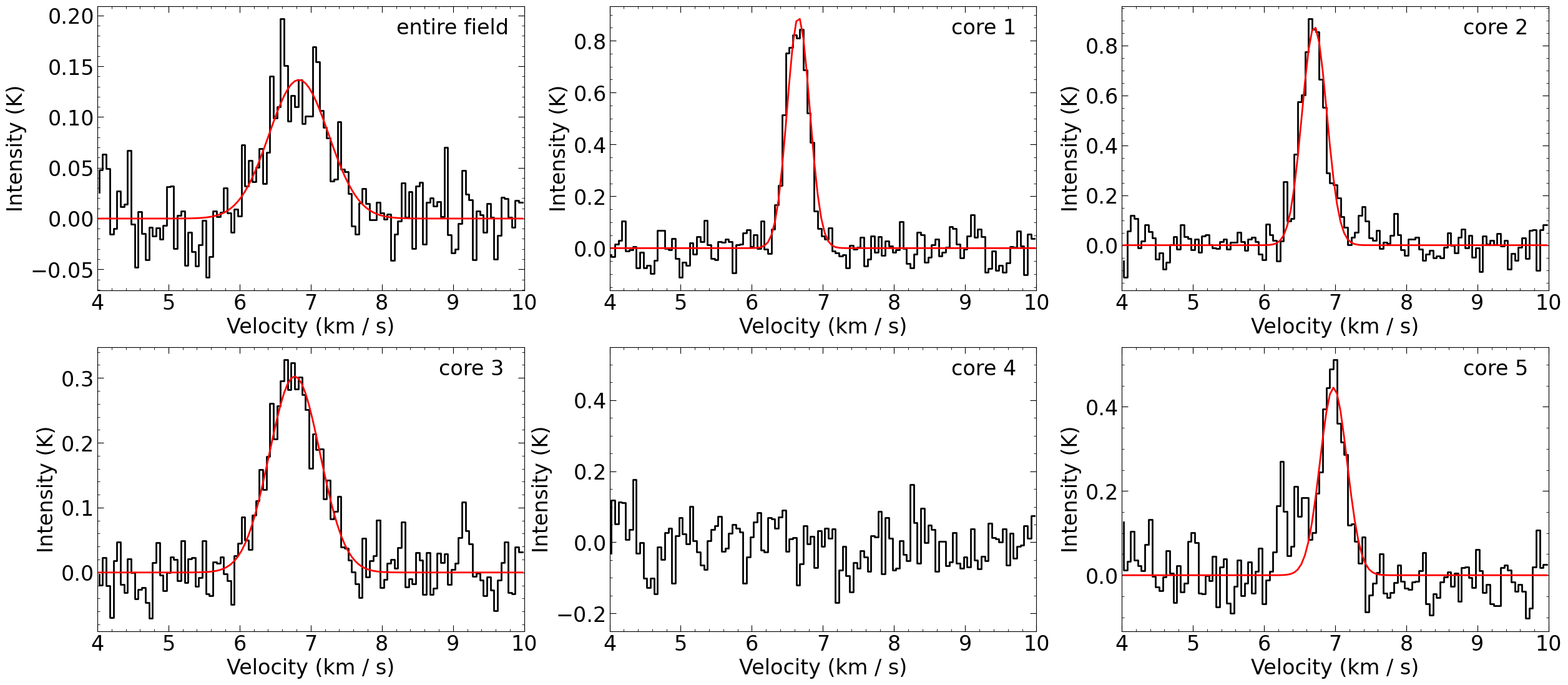}
\end{tabular}
      \caption[]{CCS spectra averaged over the entire mapping field {\bf (top left)}, at core 1  {\bf (top middle)}, at core 2  {\bf (top right)}, at core 3  {\bf (bottom left)}, at core 4  {\bf (bottom middle)}, and at core 5  {\bf (bottom right)}.}
\label{fig:spec}
\end{figure*}
The external pressure is calculated as
\begin{equation}
    P_{\rm ex} =  3 \rho_{\rm amb} \sigma_{\rm tot,amb}^2
,\end{equation}
where $\rho_{\rm amb}$ and $\sigma_{\rm tot, amb}$ are the density and 1D intrinsic velocity dispersion of ambient gas, respectively, and they are derived from the values averaged in the branch immediately below the corresponding leaf. 
The ambient density was computed under the assumption that the branch is spherical symmetry.
The core virial parameter is calculated as
\begin{equation}
\alpha_{\rm vir} =\frac{U}{W}  =\frac{3\sigma_{\rm tot}^{2}R_{\rm core}}{GM_{\rm core}}  \, .
\label{eq:virial}
\end{equation}

\begin{figure}
\includegraphics[width=0.5 \textwidth]{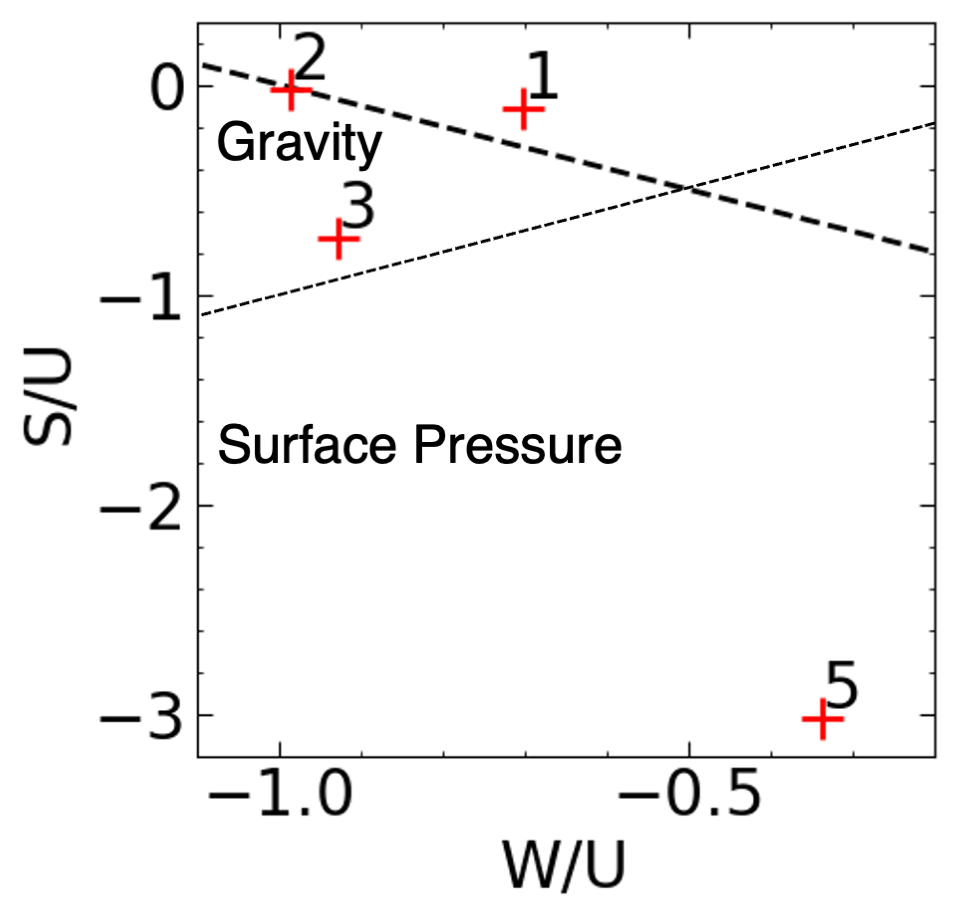}
      \caption[]{Dynamical stability of the cores based on the CCS velocity dispersion. The black dashed line represents $S+W+U = 0$.}
\label{fig:stability}
\end{figure}

Table \ref{tab:dendo_core2} summarises the velocity dispersions of the cores
and their energy ratios.
Figure \ref{fig:stability} shows the energy ratios $W/U$ versus $S/U$, where W, U, and S are the gravity term, internal kinetic and thermal energy term, and the surface term, respectively.
In this plot, the cores below the black dashed line are satisfied with the condition $S+W+U < 0$, for which the cores collapse.
The core's self-gravity is important for the area above the dotted line ($|S/U| < |W/U|$), whereas the surface pressure is dominant in the area below the dotted line.
The cores 1, 2, and 3 are located in the gravity-dominant area. They are close to the dynamical equilibrium and self-gravitating. 
For the core 5, the external pressure plays a crucial role in the core's dynamical stability, and thus the core is dynamically unstable due to the external pressure. Such a core is difficult to create from the gravitational fragmentation of the filament. The formation of unstable non-self-gravitating core (core 5) appears to agree with the collision scenario.

\section{Channel maps}

Figure \ref{fig:channelmap} shows the channel maps of HC$_3$N ($J=5-4)$ and HC$_5$N ($J=17-16$). The HC$_3$N has a stronger emission but having several overlapped hyperfine components. However, adopting the rest frequency of $J=5-4, F=5-4$ as a centre frequency, we simply integrated all the emission to construct the channel maps.
The bridge is prominent for HC$_3$N ($J=5-4)$, and structure similar to CCS was detected.
For HC$_5$N ($J=17-16)$, the strong emission is seen just at the southern edge of cores 1 and 2.

\begin{figure*}[htbp]
 \begin{center}
  \includegraphics[angle=0, width=18cm]{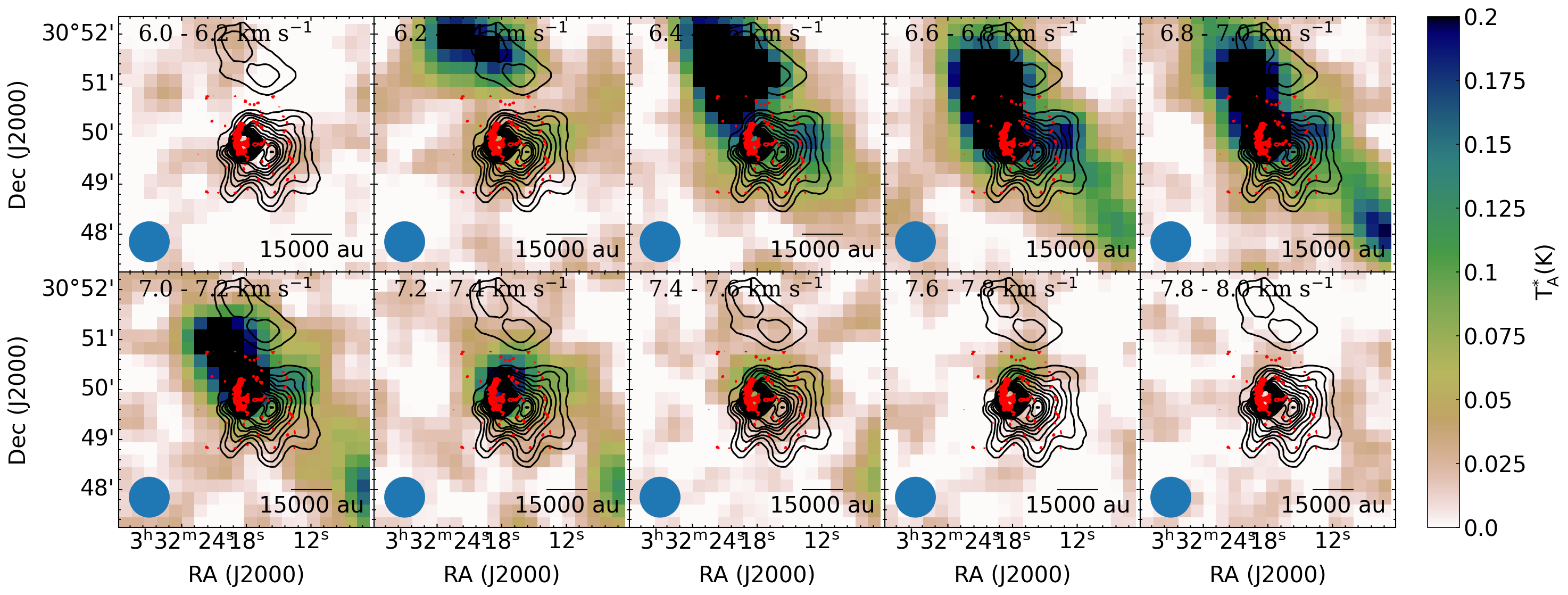}\hspace{1cm} \\
   \includegraphics[angle=0, width=18cm]{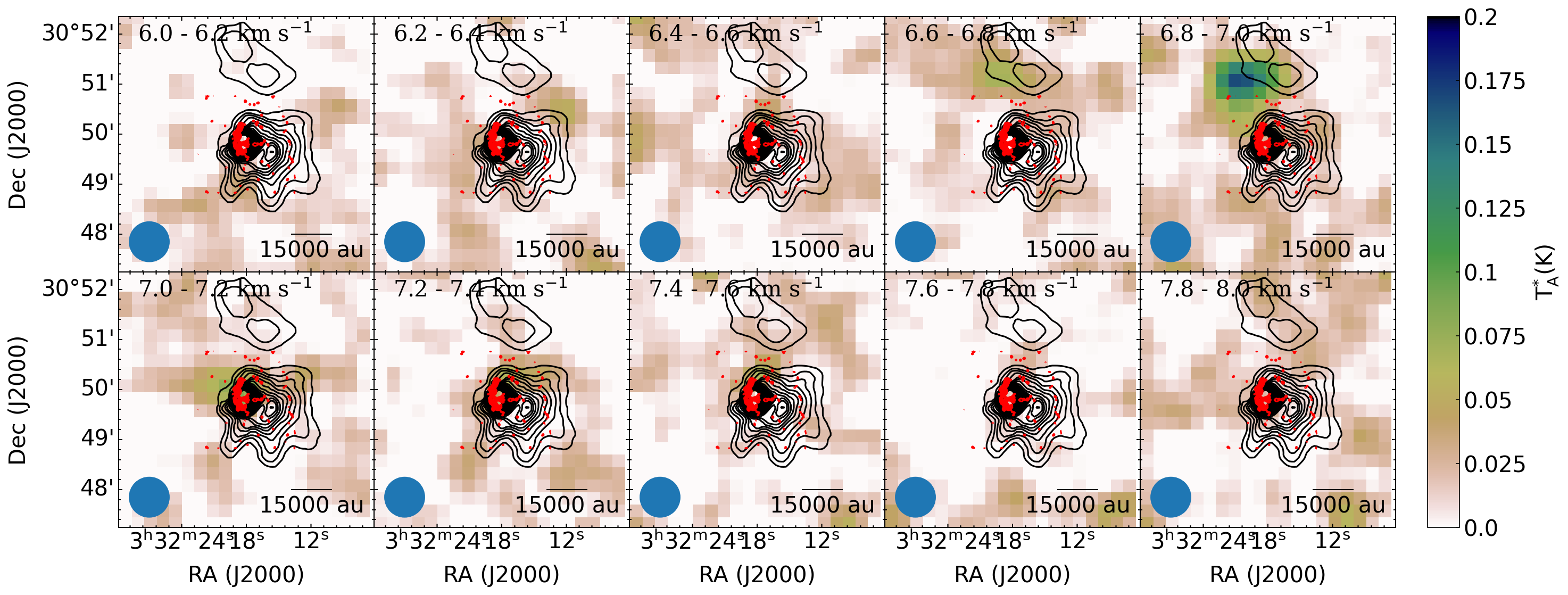}\hspace{1cm} 
 \end{center}
\caption{Colour-scale image showing channel maps of HC$_3$N ($J=5-4$) ${\bf (upper)}$ and HC$_5$N ($J=17-16$) ${\bf (lower)}$. The red contour shows the CCS ($J_N=8_7-7_6$) integrated intensity contours from NOEMA obtained by \cite{pineda20}. The black contours show {\it Herschel} column densities derived by \cite{pezzuto21}, from $1\times10^{22}$ to $10\times10^{22}$~cm$^{-2}$ in steps of $0.2\times10^{22}$~cm$^{-2}$.
Note that the HC$_3$N has overlapping hyperfine components and that we made the cube data by adopting the rest frequency indicated in Table \ref{tab:targetline}. The blue ellipse is the map effective beam size.
}
\label{fig:channelmap}
\end{figure*}

\hspace*{-1cm}
 \begin{table*}[htb]
 \footnotesize
  \caption{Dynamical properties of cores detected in the mapping area by \texttt{astrodendro}.
}
  \begin{tabular}{ccccccccc}
 \hline
 Number & $\sigma_{\rm CCS}$ & $\sigma_{\rm CCS,amb}$ & $\sigma_{\rm tot}$ & $\rho_{\rm amb}/(\mu m_H)$ & $\sigma_{\rm tot,amb}$ & $W/U$ & $S/U$ & $\alpha_{\rm vir}$ \\
  & (km s$^{-1}$) & (km s$^{-1}$) & (km s$^{-1}$) & (cm$^{-3}$) & (km s$^{-1}$) &  &  &  \\
 \hline
 1 & 0.37 & 0.41 & 0.26 & $4.53\times 10^4$& 0.28 & -0.703 & -0.111 & 1.422 \\
 2 & 0.41 & 0.41 & 0.27 & $4.53\times 10^4$ & 0.27 & -0.986 & -0.019 & 1.014 \\
 3 & 0.84 & 0.61 & 0.41 & $3.02\times 10^4$ & 0.33 & -0.928 & -0.734 & 1.077 \\
 5 & 0.45 & 0.61 & 0.28 & $2.49\times 10^4$ & 0.34 & -0.339 & -3.025 & 2.954 \\
 \hline
 \end{tabular}
\label{tab:dendo_core2}
\end{table*}

\section{Centre of gravity of the protostellar core with respect to the large filament}

The circle in Fig. \ref{fig:largemap} indicates the position of the protostar. The black square indicates the 
column density weighted position of the protostellar core. If the gas were distributed only on the plane of sky, the position would coincide with the centre of gravity. In other words, the protostar was not formed at the core centre but near the core envelope.
This implies that the core formation did not occur spontaneously.

If the curved shape of the streamer identified by \citet{pineda20} is due to the angular momentum of infalling gas, the rotational motion must be significant on the scale of $10^3$ au, leading to the formation of a large ($\sim 10^3$ au) rotating envelope or disk. However, such a large circumstellar structure was not detected in the observations by \citet{tobin16}. Thus, at this scale, the rotational motion would still be minor and unlikely to create the elongated, curved structure seen in Fig. \ref{fig:CCSchannelmaps}. For example, \citet{yano24}'s core collision simulation showed that the streamer formed by the infalling gas from the second core is smoothly connected to a central, rotating disk-like envelope. Such a structure is a plausible outcome of the infalling gas scenario. Since the rotating disk is small for this protostar, the curved structure, if formed by the infalling gas, must almost point directly to the protostar.
In Fig. \ref{fig:largemap} we have also indicated the position of the filament shown in Fig. \ref{fig:model}.
\begin{figure}
\begin{tabular}{c}
\includegraphics[width=12.cm]{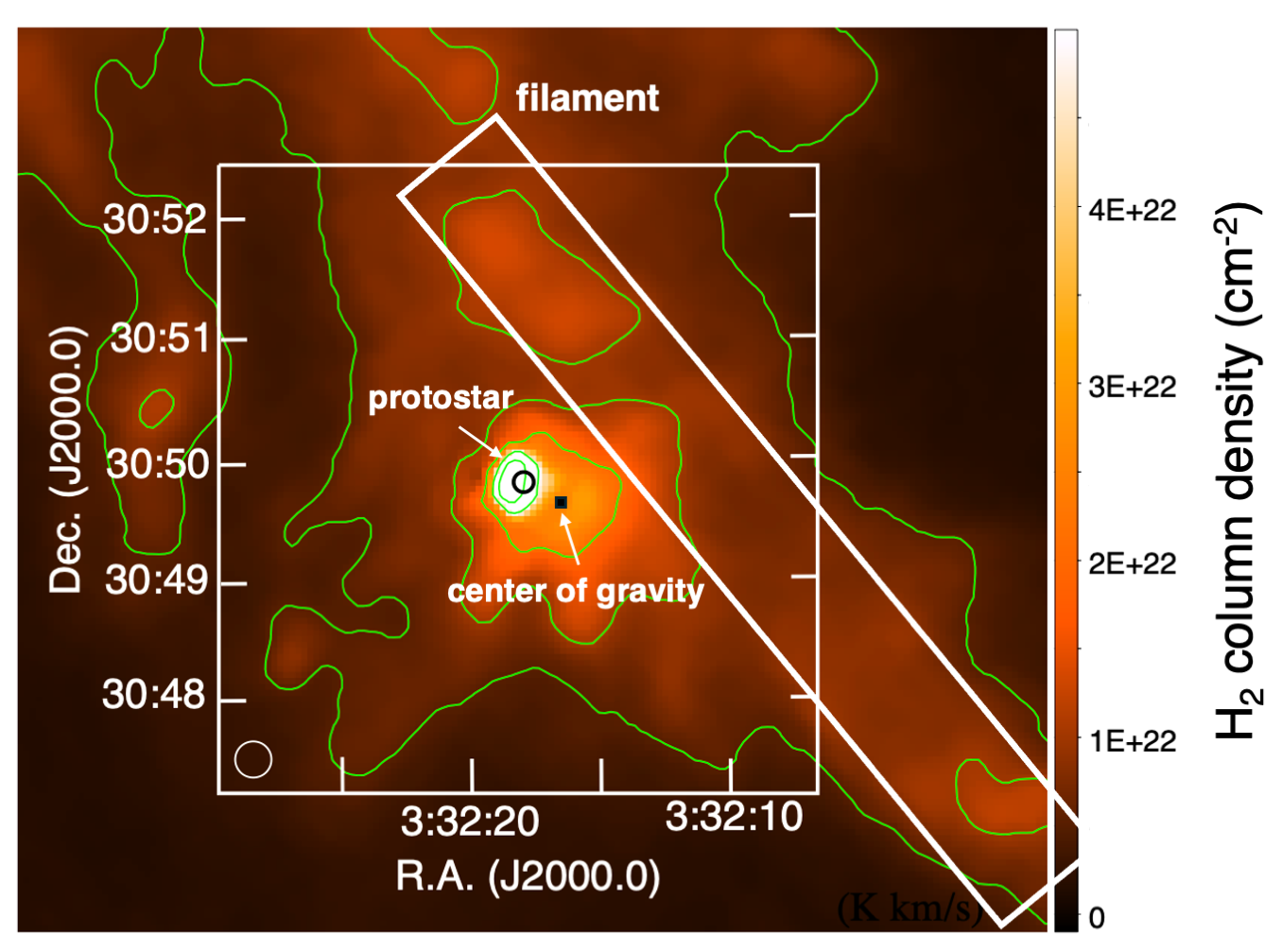}
\end{tabular}
      \caption[]{{\it Herschel} H$_2$ column density image. The white rectangle indicates the filament that collided with the dense core in our CFC model. The white box with the coordinates is the observation box.
      The circle at the bottom-left corner represents the beam size of the {\it Herschel} map. The small circle near the centre is the position of the protostar.
      The colour represents the H$_2$ column density in cm$^{-2}$. 
      The green contours are drawn at the H$_2$ column density levels of 5 $\times 10^{21}$, 2 $\times 10^{22}$, 3 $\times 10^{22}$, 5 $\times 10^{22}$, and 1 $\times 10^{23}$ cm$^{-3}$.}
\label{fig:largemap}
\end{figure}


\end{appendix}
\end{document}